\newif\ifReferee
\Refereefalse
\ifReferee
\documentclass[referee,usenatbib,usegraphicx,onecolumn]{mn2e}
\else
\documentclass[usegraphicx,useAMS,usenatbib,onecolumn]{mn2e}
\fi
\IfFileExists{times.sty}{\usepackage{times}}{}
\IfFileExists{1txfonts.sty}{\usepackage{txfonts}}{
\ifx\undefined\iint
\newcommand{\iint}{\int\!\!\!\int}
\fi
}
\usepackage{color}
\ifx\undefined\bmath
\newcommand{\bmath}[1]{\mbox{\boldmath$#1$}}
\fi
\ifx\undefined\mathbfss
\newcommand{\mathbfss}[1]{\mbox{\bf\textsf{#1}}}
\fi
\ifx\undefined\bm
\newcommand{\bm}[1]{\bmath{#1}}
\fi
\ifx\undefined\rmn
\newcommand{\rmn}[1]{\mathrm{#1}}
\fi
\begin{document}
\title{
Finite source effect on the polarization degree induced by a single
microlens 
}
\author[Hiroshi Yoshida]{
Hiroshi Yoshida$^1$\thanks{E-mail: yoshidah@fmu.ac.jp}\\ \\
$^1$ Department of Physics, Fukushima Medical University, Fukushima-City
960-1295, Japan
}
\maketitle
\begin{abstract}
 We investigate the effect of a single microlens on Stokes parameters. 
 Semi-analytical formulae of the microlensed Stokes parameters are
 derived. The formulae not only reduce the double integrals in the
 estimations of those quantities but can also be approximated to a
 useful form in the bypass case. By using our formulation, we show that 
 a combination of polarimetric data with photometric data enables us to
 estimate not only the finite source effect but also the direction of
 the microlens motion.
\end{abstract}
\section{Introduction}
The microlens effect is a phenomenon in which the flux emitted from a
source (star) varies with time due to the gravitational lens effect produced
by a compact object in the foreground, such as a star or a
planet. \cite{pacz86} suggested that microlensing events could be 
detected by a successive observation of stars in LMC, SMC or our
Galactic bulge, and since 1993 \citep{alco93,aubo93,udal93} many
observational groups have reported a substantial number of microlensing
events. According to \citeauthor{pacz86}, the observations of microlensing
events provide information on the mass distribution of invisible objects
in galaxies; further, they impose some restrictions on the dark matter. 
\par
The observation of a microlensing event provides a light curve of the
source, which is a fundamental observable in this event. From analysis of
the light curve, it is important to estimate the microlens parameters
such as the mass $M$ of the microlens object, the distance parameter
$\zeta$($=D_\rmn{OL}/D_\rmn{OS}$, where $D_\rmn{OL}, D_\rmn{OS}$ are
the distances from the microlens and from the source to the observer,
respectively) and relative velocity $\bmath{V}$ of 
the microlens orthogonal to the line of sight. 
However, it is usually very difficult to obtain the value of each
parameter because these parameters are coupled with each other in
observed quantities, e.g. the event duration 
$\hat{t}=2t_\rmn{E}=2r_\rmn{E}/|\bmath{V}|$ and the normalized minimum impact
parameter $u_0=b/r_\rmn{E}$, where
$r_\rmn{E}=(4GMD_\rmn{OS}\zeta(1-\zeta)/c^2)^{1/2}$ is the Einstein ring 
radius on the lens plane. 
\par
In order to decouple the degeneracy in the microlens parameters, several
authors have proposed useful effects associated with microlens events,
namely, the parallax effect on the light curve \citep{goul92}, and the
accurate astrometry of the lensed source
\citep*{hog95,miya95,walk95}. The former effect enables us to measure the
Einstein ring radius projected on the observer plane (referred to as the
reduced Einstein ring radius
$\tilde{r}_\rmn{E}=r_\rmn{E}/(1-\zeta)$). This effect has been observed
in a large number of events
\citep*{macho95,Mao99,macho2000,Bond01,Mao02,Smith02, 
Smith03a,Afonso03,Bennet03,Smith03b}. The latter effect provides the angular
Einstein ring radius  $\theta_\rmn{E}(=r_\rmn{E}/D_\rmn{OL})$ through
the observation of the fine motion of the lensed source. In order to
detect this effect, we need the $\mu$as level accuracy in the
astrometry. Hence, some satellite projects are being planned for
carrying out such observations.These two effects are 
expected to impose restrictions on the determination of the microlens
parameters. In particular, the detection of both the effects in the same
event enables us to decouple the degeneracy in the parameters 
$M, \zeta$ and $|\bmath{V}|$ as  $M=c^2\tilde{r}_\rmn{E}\theta_\rmn{E}/4G,\ 
\zeta=\tilde{r}_\rmn{E}/(D_\rmn{OS}\theta_\rmn{E}+\tilde{r}_\rmn{E})$ 
and $|\bmath{V}|=\zeta D_\rmn{OS}\theta_\rmn{E}/t_\rmn{E}$, respectively.
\par
To date, certain events that involved the estimation of the microlens
mass without the high-precision astrometry have been reported
\citep{Smith03b,Yoo04,Park04,Ghosh04,Jiang04}.  
In four of these events, the authors 
(\citeauthor{Smith03b,Yoo04,Park04,Jiang04})
 not only detected the parallax
effect but also estimated the angular Einstein radius by combining an
empirical relation between the angular size $\theta_*$ and the
colour-magnitude of the source given by \cite{vanBell99} with the finite
source effect $\rho_*=\theta_*/\theta_\rmn{E}$ on the light curve
\citep{witt94,nemi94}. The finite source effect is relatively easy to be
detected in the transit case, where the lens transverses the face of the
source projected on the lens plane; this is because the light curve of
an extended source is significantly different from that of a point
source when the normalized impact parameter $u$ is such that $u\la\rho_*$ 
\citep[e.g.][]{macho97b,Soszy01}. The above events that are successful
in  decoupling the degeneracy in the parameters are examples in the case.
On the other hand, in the bypass case, where the lens never transverses
the face of the source,  it is difficult to detect the effects 
because we can fit the light curve using appropriate parameters in a
point source model \citep{goul94}. \cite{welch95} found
that in the bypass case the finite source effect is not detectable by a
single-band observation but can be detected by a two-band observation,
especially in the optical and infrared bands.
\par
Recently, \citet*{simm95a} and \citet*{simm95b} pointed out that the
finite source 
effect also appears in another aspect of the microlens effect. They
found that the Chandrasekhar effect \citep{chan60} is enhanced by the
microlens effect and that this enhancement causes the polarization degree
of the source to vary with the motion of the microlens. They also found that
the polarization degree has a double peak in the transit case, while it
has a single peak in the bypass case. They presented a numerical
calculation of the Stokes parameters with the microlens effect; however,
they did not present the analytical formula, although the size
dependence of the polarization degree helps us to decouple the
degeneracy in the microlens parameters. Before \citeauthor{simm95a},
\cite{schn87} also discussed the gravitational lens effect on the
polarization degree as a tool for identifying a lensed supernova. In
their paper, by using a simple approximation of the magnification
factor\footnote{%%% footnote 1
According to \cite*{Sch92}, we use the term
`magnification factor' instead of `amplification factor'.
} for a
point lens, they presented a semi-analytical formula of the polarization
degree and the magnification factor to demonstrate that the size
expansion during the supernova explosion yields variations in these
quantities. However, they did not discuss the benefits from the
polarimetry of the microlensed source at all. \par
In this paper, we focus on the finite source effect
on the Stokes parameters and the polarization degree of the microlensed
star. Further, we will also discuss the possibility that the polarimetric
observation of the lensed source decouples the degeneracy in the
microlens parameters. Throughout this paper, we consider a single microlens.
The outline of this paper is as follows. In Section 2, we provide a
brief review of the microlens effect. In Section 3, we present
semi-analytical formulae of the Stokes parameters to discuss the finite
source effect on the polarization degree. Then, we provide an
approximation of the formula with regard to the bypass case. In Section
4, we consider the extraction of information concerning the 
finite source effect from the polarimetric observation of the
microlensed source. Finally, in Section 5, we discuss some of the 
possibilities of detecting the polarization induced by the microlens
effect. 
\section{Microlens effect on the light curve}
When light rays from a distant source pass near a compact object
(microlens) with mass $M$, they are deflected by an angle $4GM/c^2b$,
where $b$ is an impact parameter. Let $D_{\mathrm{OS}}, D_{\mathrm{OL}}$
be the distances from the source and from the microlens to the observer,
respectively. Let $\bm{\theta}_{\mathrm{S}}$ be the angular
position of the source that would be observed without the microlens. In
this situation, the lensed light (hereafter referred to as `image')  
is observed at the angular position $\bm{\theta}$, which is obtained as
a solution of the following lens equation:
\begin{equation}
\bm{\theta}_{\mathrm{S}}=\bm{\theta}-{\theta_\rmn{E}^2\over 
\theta^2}\bm{\theta},
\label{eq:lenseq} 
\end{equation}
where 
$\theta_\rmn{E}=\sqrt{4GM(1-\zeta)/\zeta c^2D_\rmn{OS}}$ is the angular
Einstein ring radius. The distance parameter $\zeta$ is given by
$D_\rmn{OL}/D_\rmn{OS}$. In equation
(\ref{eq:lenseq}), let us introduce the dimensionless variables 
$\bmath{u}\equiv \bmath{\theta}_\rmn{S}/\theta_\rmn{E}$ and
$\bmath{x}\equiv\bmath{\theta}/\theta_\rmn{E}$ and rewrite the lens
equation as follows: 
\begin{equation}
\bmath{u}=\bmath{x}-{\bmath{x}\over |\bmath{x}|^2}. \label{eq:lenseqx}
\end{equation}
\begin{figure}
\begin{center}
\includegraphics[width=0.3\linewidth]{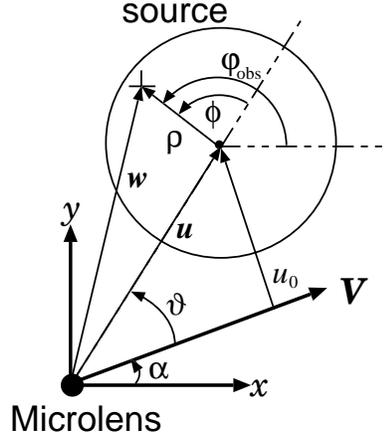}
\end{center}
\caption{Geometry of microlens and source:
 $\bm{u}$ and $\bm{w}$ are the vectors to the
 centre and to a point marked as `+' at $(\rho,\phi)$ of the source
 projected on the lens plane from the microlens,
 respectively. $\vartheta$ and  $\alpha$ are the angles between $\bm{V}$
 and $\bm{u}$ and between $\bm{V}$ and the $x$--axis defined by an
 observer, respectively. Therefore,
 $\varphi_\rmn{obs}=\phi+\vartheta+\alpha$ is the angle between the
 $x$--axis and the line joining the marked point and the centre. $u_0$ is
 the minimum impact parameter. 
}%%% Figure 1
\label{fig:geometry}
\end{figure}
In general, the equation has two solutions, $\bmath{x}_\pm$, and
they give the image positions of the source. One of the images is
gravitationally magnified by factor $A_+$, while the other is magnified
by factor $A_-$ (demagnified):
\begin{equation}
A_\pm(u)=\frac{1}{2}\left({u^2+2\over
		  u\sqrt{u^2+4}}\pm1\right).\label{eq:Imagemag} 
\end{equation}
However, we cannot resolve each image because the typical separation
angle between the images is 
$\sim\theta_\rmn{E}\la0.1-1(M/M_\odot)^{1/2}\rmn{mas}$.
Hence, we observe them as a single image with the following
magnification factor: 
\begin{equation}
A_\rmn{p}(u)=A_+(u)+A_-(u)={u^2+2\over u\sqrt{u^2+4}}. \label{eq:magpf}
\end{equation}
\par
Because the microlens moves with a velocity $\bmath{V}$ relative to the
line of sight, the normalized distance $u$ from the
microlens to the source varies with time as
$u=\{u_0^2+\tau^2\}^{1/2}$, where $u_0$ is the normalized minimum impact
parameter of the microlens, $\tau=(t-t_0)/t_\rmn{E}$ is the 
time normalized by the time-scale of the microlensing event
($t_\rmn{E}=D_\rmn{OL}\theta_\rmn{E}/\vert\bmath{V}\vert$) and $t_0$ is 
the time at which the magnification factor has a maximum
value. Consequently, the microlens motion yields the variation of the
magnification factor.
\par
By taking into account the finite size of the source, some authors
pointed out a remarkable difference between the light curves of the
point source (the standard microlens) model and of the extended source
model \citep{witt94,nemi94}. This effect is termed the finite source
effect.  In general, the lensed flux is given by
\ifReferee
\begin{eqnarray}
S_I(\rho_*|u)&=&\int_0^{\rho_*}\theta_\rmn{E}^2\rho\rmn{d}\rho 
\int_{-\pi}^\pi I(\theta_\rmn{E}\rho,\phi)A_\rmn{p}(w(u,\rho,\phi))\rmn{d}\phi
\equiv A(\rho_*|u)S_{I0},\label{eq:obsFlux}\\
S_{I0}&=&\int_0^{\rho_*}\theta_\rmn{E}^2\rho\rmn{d}\rho\int_{-\pi}^\pi
I(\theta_\rmn{E}\rho,\phi)\rmn{d}\phi,\nonumber
\end{eqnarray}
\else
\begin{equation}
S_I(\rho_*|u)=\int_0^{\rho_*}\theta_\rmn{E}^2\rho\rmn{d}\rho 
\int_{-\pi}^\pi I(\theta_\rmn{E}\rho,\phi)A_\rmn{p}(w(u,\rho,\phi))\rmn{d}\phi
\equiv A(\rho_*|u)S_{I0},\quad
S_{I0}=\int_0^{\rho_*}\theta_\rmn{E}^2\rho\rmn{d}\rho\int_{-\pi}^\pi
I(\theta_\rmn{E}\rho,\phi)\rmn{d}\phi,\label{eq:obsFlux}
\end{equation}
\fi
where $I(r,\phi)$ is the surface brightness and  $A(\rho_*|u)$ denotes
the total magnification factor.
In the above equations, $\rho$ denotes the normalized radial
distance to a point on the lens plane from the centre of the projected
source, while $\phi$ denotes the angle between $\bmath{u}$ and the 
line joining the  point and the centre (Fig. {\ref{fig:geometry}}).  
Furthermore, 
$\rho_*=\theta_*/\theta_\rmn{E}$ ($\theta_*$: the angular radius of 
the source) is the normalized radius of the source,  and $w$ is the
normalized distance from the microlens to the point, given by  
$w(u,\rho,\phi)=\sqrt{u^2+2u\rho\cos\phi+\rho^2}$.
\par
The total lensed flux from the extended source with axial symmetry
in the surface brightness, $I(r,\phi)=I(\theta_\rmn{E}\rho)$, is given by
\begin{equation}
S_I(\rho_*|u)
=4\theta_\rmn{E}^2\int_0^{\rho_*}\rho\rmn{d}\rho
I(\theta_\rmn{E}\rho)
\left[
{(u-\rho)^2\Pi(\pi/2;k,n)+2F(\pi/2;k)\over
 (u+\rho)\sqrt{(u-\rho)^2+4}}\right],\label{eq:totFlux}
\end{equation}
where $F(\pi/2;k)$ and $\Pi(\pi/2;k,n)$
are the complete elliptical integrals of the first and of the third
kinds, respectively \citep[][ \ see also Appendix A]{grad96}. Further,
\begin{equation}
n={4u\rho\over(u+\rho)^2},\quad k=\sqrt{4n\over(u-\rho)^2+4},
\label{eq:n_k}
\end{equation}
were introduced by \citeauthor{witt94}, where the total magnification factor
of the flux from the source with a constant surface brightness was
estimated. An equivalent expression of equation (\ref{eq:totFlux}) was 
also presented by \cite{witt95}, which enables us to numerically
calculate the total fluxes in various bands. Using this
expression, we can obtain the total magnification for the source with
the limb-darkening surface brightness
\begin{equation}
I(\theta_\rmn{E}\rho)=
I_0\left[1-c_1\left(1-\sqrt{1-{\rho^2\over\rho_*^2}}\right)\right] 
\label{eq:surfbr}
\end{equation}
as follows:
\ifReferee
\begin{eqnarray}
A(\rho_*|u)&=&{4\over\pi\rho_*^2(1-c_1/3)}
\int_0^{\rho_*}\rmn{d}\rho \rho\left[
1-c_1\left(1-\sqrt{1-{\rho^2\over\rho_*^2}}\right)
\right]\nonumber\\&&\hspace{50mm}\times
{(u-\rho)^2\Pi(\pi/2;k,n)+2F(\pi/2;k)\over(u+\rho)\sqrt{(u-\rho)^2+4}},
\label{eq:extotmag}
\end{eqnarray}
\else
\begin{eqnarray}
A(\rho_*|u)={4\over\pi\rho_*^2(1-c_1/3)}
\int_0^{\rho_*}\rmn{d}\rho \rho\left[
1-c_1\left(1-\sqrt{1-{\rho^2\over\rho_*^2}}\right)
\right]
{(u-\rho)^2\Pi(\pi/2;k,n)+2F(\pi/2;k)\over(u+\rho)\sqrt{(u-\rho)^2+4}},
\label{eq:extotmag}
\end{eqnarray}
\fi
where $c_1$ is a parameter that depends on the wavelength of the light.
\par
\begin{figure*}
\begin{center}
\rotatebox{-90}{\includegraphics[width=0.45\linewidth,clip]{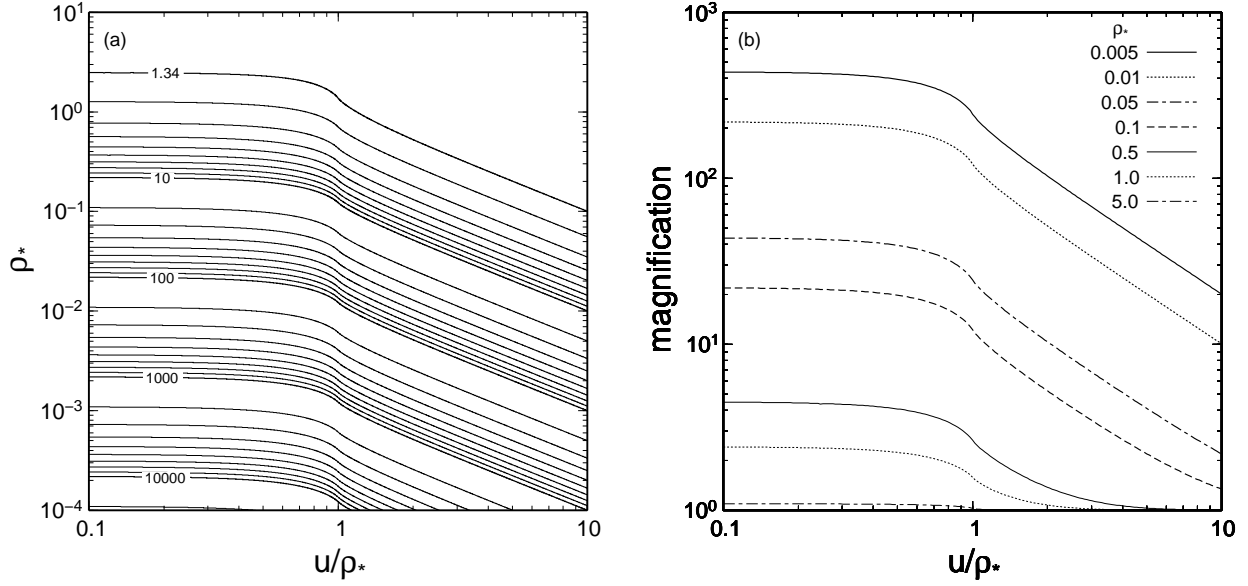}}
\caption{Magnification factor: contours of the magnification factor (a) 
 and the $u/\rho_*$ dependency of the magnification factor (b). In our
 estimation, the linear limb-darkening model
 $I(\theta_\rmn{E}\rho)=I_0\{1-c_1[1-(1-\rho^2/\rho_*^2)^{1/2}]\}$ is
 adopted, where  $c_1=0.64$.
}%%% Figure 2
\label{fig:lightcurve2}
\end{center}
\end{figure*}
In Fig. \ref{fig:lightcurve2}, we present the contours of $A(\rho_*|u)$ on
the $u/\rho_*$-$\rho_*$ plane in panel (a) and the $u/\rho_*$ dependence of
$A(\rho_*|u)$ in panel (b). Both panels clearly show that the behaviour of
$A(\rho_*|u)$ inside the source is very different from that outside the
source. Outside the source, the contours (Fig. \ref{fig:lightcurve2}-a),
particularly, for low $\rho_*$, are almost straight lines with a slope of
$\rho_*/u$, which shows that the magnification factors depend only on $u$.
On the other hand, the contours inside the source are rather flat, which
means that the factors inside the source have little dependence on $u$.
\par
\section{Microlens effect on the Stokes parameters} 
\subsection{Stokes parameters}
\cite{chan60} numerically calculated the effect of electron
scattering in the atmosphere of a star on the light emitted from the
star and found that it leads to polarization (the \citeauthor{chan60}
effect). 
According to \cite{schn87}, we can assume that the Stokes
parameters $Q,U,V$ at each point $(\rho,\phi)$ on the surface of the
star with axial symmetry are expressed as follows: 
\begin{equation}
Q(\theta_\rmn{E}\rho)
=I_0c_2\left(1-\sqrt{1-{\rho^2\over\rho_*^2}}\right),\ U=V=0.
\label{eq:IQUV}
\end{equation}
\par
Normally, we cannot resolve the light from each point on a star except
for the Sun. We only observe the total flux that is obtained by
integrating the lights over the surface of the star. In this case, the
total Stokes parameters, $S_Q, S_U$ and $S_V$, are also defined as those
integrated over the surface. In integrating the Stokes parameters over
the surface, it must be done based on the same reference frame. In this
regard, we must fix $x$- and $y$- axes such as in the
West and North directions, respectively. As seen in
Fig. {\ref{fig:geometry}}, we define the angle between $x$-axis defined
by the observer and the line joining the centre of the star and the
point $(\rho, \phi)$ as $\varphi_\rmn{obs}$. We thus define the total
Stokes parameters as follows: 
\begin{equation}
\left(
\begin{array}{c}
S_Q\\
S_U\\
S_V\end{array}
\right)
=\int_0^{\rho_*}\theta_\rmn{E}^2\rho\rmn{d}\rho Q(\theta_\rmn{E}\rho)
\int_{-\pi}^\pi\rmn{d}\varphi_\rmn{obs}\left(
\begin{array}{c}
\cos2\varphi_\rmn{obs}\\
-\sin2\varphi_\rmn{obs}\\
0\end{array}
\right).
\end{equation}
In this case, we can easily deduce that $S_Q=S_U=S_V=0$ because of the
axial symmetry. Therefore, the light from the usual circular star is
unpolarized. Specifically, we cannot detect the Chandrasekhar effect in a star
without any symmetry breaking. In fact, \cite{kemp83} detected this 
effect during an eclipse in a binary system. 
\par
\subsection{Microlensing on polarization}
\cite{simm95a} pointed out that due to the symmetry breaking by the
differential gravitational magnification, the cancellation of 
polarization between symmetrical points on the surface is
incomplete, and the flux is totally polarized as a result. Moreover,
the magnification of light from each part of the surface also varies
with the microlens motion. Consequently, the polarization varies with time.
\cite{simm95b} also numerically calculated the polarization degree and
found that the time profile of the polarization has a single peak in the
bypass case, while it has double peaks in the transit case. 
Unfortunately, it is difficult to understand the finite source effect on
polarization, especially the size dependence of polarization, only on
the basis of the numerical calculation. 
\par
According to \cite{simm95b}, the lensed Stokes parameters are given by
\begin{equation}
\left(\begin{array}{c}
S_Q\\
S_U
\end{array}
\right)=\int_0^{\rho_*}\theta_\rmn{E}^2\rho\rmn{d}\rho
Q(\theta_\rmn{E}\rho) \int_{-\pi}^\pi\rmn{d}\varphi_\rmn{obs}
\left(\begin{array}{c}
\cos2\varphi_\rmn{obs}\\
-\sin2\varphi_\rmn{obs}
\end{array}
\right)A_\rmn{p}(w(u,\rho,\phi)).
\label{eq:ASQUa}
\end{equation}
The relation between $\varphi_\rmn{obs}$ and $\phi$ is
$\varphi_\rmn{obs}=\phi+\vartheta+\alpha$, where 
$\vartheta=-\tan^{-1}(u_0/\tau)$ and $\alpha$ is the angle between the
direction of the microlens motion $\bmath{V}$ and the $x$-axis (see
Fig. {\ref{fig:geometry}}). 
From this relation, we can integrate the inner part of equation
(\ref{eq:ASQUa}) with respect to $\phi$ and express the Stokes parameters
as follows: 
\ifReferee
\begin{eqnarray}
\left(\begin{array}{c}
S_Q\\
S_U
\end{array}
\right)&=&D(\rho_*|u)\left(\begin{array}{c}
\cos2(\vartheta+\alpha)\\ -\sin2(\vartheta+\alpha)\end{array}\right),
%\hbox{\qquad\hspace{37mm}}
\label{eq:ASQU}\\
D(\rho_*|u)&=&4I_0\theta_\rmn{E}^2c_2\int_0^{\rho_*}\rmn{d}\rho\left(1-\sqrt{1-{\rho^2\over\rho_*^2}}\right)\left[
{(u-\rho)^2\over u^2\rho(u+\rho)\sqrt{(u-\rho)^2+4}}
\Pi({\pi/2};k,n)
\right.\nonumber\\&&\hspace{6mm}
+{(u^2-\rho^2)^2(u^2+\rho^2)+6(u^4+\rho^4)+4u^2\rho^2
\over8u^2\rho(u+\rho)\sqrt{(u-\rho)^2+4}}F({\pi/2};k)
\nonumber\\&&\hspace{10mm}
\left.-{(u+\rho)(u^2+\rho^2+2)\sqrt{(u-\rho)^2+4}\over8u^2\rho}
E({\pi/2};k)\right],
\label{eq:ASU}
\end{eqnarray} 
\else
\begin{eqnarray}
\left(\begin{array}{c}
S_Q\\
S_U
\end{array}
\right)&=&D(\rho_*|u)\left(\begin{array}{c}
\cos2(\vartheta+\alpha)\\ -\sin2(\vartheta+\alpha)\end{array}\right),
%\hbox{\qquad\hspace{37mm}}
\label{eq:ASQU}\\
D(\rho_*|u)&=&4I_0\theta_\rmn{E}^2c_2\int_0^{\rho_*}\rmn{d}\rho\left(1-\sqrt{1-{\rho^2\over\rho_*^2}}\right)\left[
{(u-\rho)^2\over u^2\rho(u+\rho)\sqrt{(u-\rho)^2+4}}
\Pi({\pi/2};k,n)
\right.\nonumber\\&&\hspace{2mm}
+{(u^2-\rho^2)^2(u^2+\rho^2)+6(u^4+\rho^4)+4u^2\rho^2
\over8u^2\rho(u+\rho)\sqrt{(u-\rho)^2+4}}F({\pi/2};k)
%\nonumber\\&&\hspace{10mm}
\left.-{(u+\rho)(u^2+\rho^2+2)\sqrt{(u-\rho)^2+4}\over8u^2\rho}
E({\pi/2};k)\right],
\label{eq:ASU}
\end{eqnarray} 
\fi
where $E(\pi/2;k)$ is the complete elliptical integral of the second
kind; $k$ and $n$ are given by equation (\ref{eq:n_k}). 
Hence, we obtain the polarization degree as follows: 
\begin{equation}
P(\rho_*|u)={\sqrt{S_Q^2+S_U^2}\over S_I(\rho_*|u)}={D(\rho_*|u)\over S_I(\rho_*|u)}.
\label{eq:p_def}
\end{equation}
\begin{figure*}
\begin{center}
\rotatebox{-90}
{\includegraphics[width=0.45\linewidth]{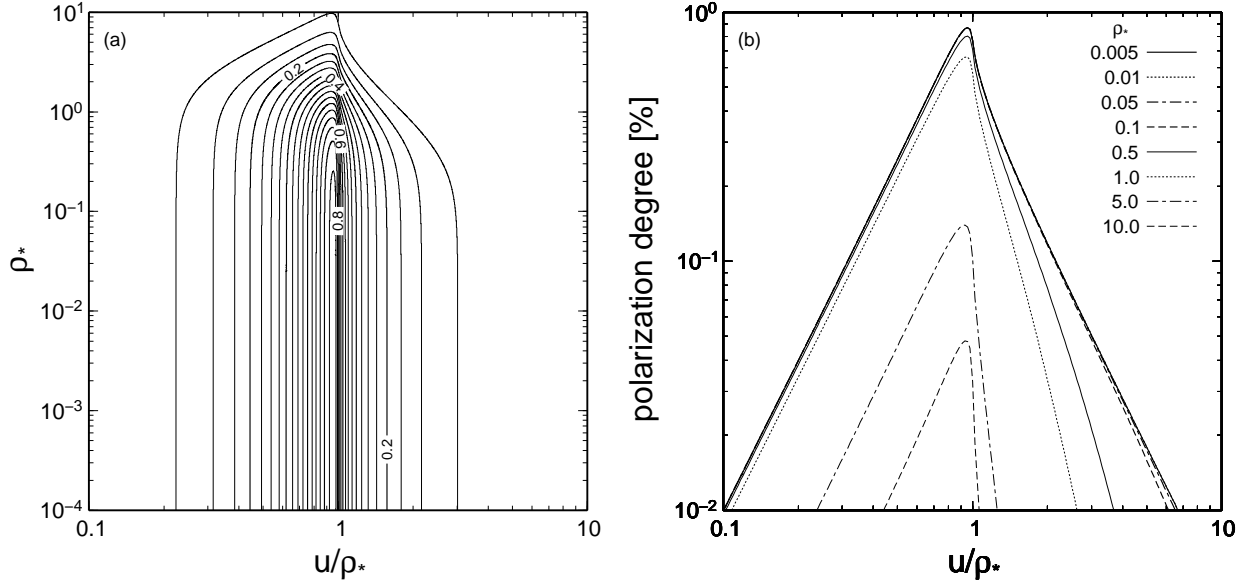}}
\caption{Polarization degree: Contours of the polarization (a) and the
 $u/\rho_*$ dependency of the polarization degree (b). In this
 estimation, the limb-darkening model
 $Q(\theta_\rmn{E}\rho)=I_0c_2[1-(1-\rho^2/\rho_*^2)^{1/2}]$ 
 is adopted, where  $c_2=0.04$. In panel (a), the contour  spacing is
 0.05 per cent.
}%%% Figure 3
\label{fig:polarization2} 
\end{center}
\end{figure*}
Using equations (\ref{eq:totFlux}) and (\ref{eq:ASU}), we can reduce the
estimation of the polarization degree from a double integral to a single
integral in equations (\ref{eq:obsFlux}) and (\ref{eq:ASQUa}). In the
numerical estimation, we can use some subroutines (e.g. GNU Scientific
Library\footnote{%%% footnote 2
\texttt{http://www.gnu.org/software/gsl/index.html}}) for 
the complete elliptical integrals in the these equations.
\par
Fig. \ref{fig:polarization2} shows the contours of
polarization degree on the $u/\rho_*$-$\rho_*$ plane in panel (a) and
the $u/\rho_*$-dependence on the polarization degree in panel (b). From
panel (a), it is clear that for $\rho_*\la0.05$, the contours are
parallel to the $\rho_*$-axis at least up to $u/\rho_*\la10$. This means
that the polarization degree depends only on $u/\rho_*$ for $\rho_*\la0.05$ and
$u/\rho_*\la5$, which is also shown in panel (b). As mentioned in
\citeauthor{schn87}, we also confirm that the polarization degree has the
maximum value $P_\rmn{max}\approx0.875~ \rmn{per~ cent}$ at
$u/\rho_*\approx0.96$ for low 
$\rho_*$. On the other hand, for large $\rho_*$, it is apparent from panel
(b) that the polarization degree strongly depends on the star's
radius. From panel (b), furthermore, we observe that while for $u/\rho_*<1$,
the polarization is almost proportional to $(u/\rho_*)^2$, as shown by
\citeauthor{schn87}, whereas for $u/\rho_*>1$, the behaviour of the
polarization is not as simple as argued by them. This difference arises
due to the fact that \citeauthor{schn87} adopted an approximation to the
magnification factor for the point lens, while we refrained from doing
this in this stage. Our expression (eq. [\ref{eq:ASU}]) is the exact
form for the single microlens. 
\subsection{Polarization degree in the bypass case}
The use of equations (\ref{eq:totFlux}) and (\ref{eq:ASQU}), besides
reducing the double integral in equations (\ref{eq:obsFlux}) and
(\ref{eq:ASQUa}), is also advantageous in obtaining an
approximation of the polarization degree in the bypass case.  
Hence, in this subsection, we focus on the polarization
degree in the bypass case. By using equations (A1)-(A3) given in
the Appendix of \cite{witt94}, we can expand the complete elliptic integrals
with respect to $k$ and $n$ in the bypass case ($\rho_*<u_0$ and
$\rho_*<1$).
\footnote{%%% footnote 3
In the microlensing events reported so far that exhibit the finite
source effect (transit cases), the assumption $\rho_*<1$ is a good
approximation \citep{macho97b}.} 
We can thus approximate formulae (\ref{eq:totFlux}) and
(\ref{eq:ASU}) as follows:  
\begin{eqnarray}
S_I(\rho_*|u)&=&A_\rmn{p}(u)S_{I0}\left[1+{60-28c_1\over15-5c_1}
{u^2+1\over\left(u^2+2\right)\left(u^2+4\right)^2}
\left({\rho_*\over u}\right)^2\right],
\label{eq:SI_lu0}\\
D(\rho_*|u)&=&
{7c_2\over5(1-c_1/3)}
{A_\rmn{p}(u)S_{I0}\over \left(u^2+4\right)^2}\left({\rho_*\over u}\right)^2
\left[
1+{304\left(u^6+3u^4+6u^2+5\right)\over147\left(u^2+2\right)
\left(u^2+4\right)^2}\left({\rho_*\over u}\right)^2\right].
\label{eq:D0_lu0}
\end{eqnarray}
While the finite source effect $\rho_*$ is a secondary effect in the
magnification factor, it is the primary effect in the polarization degree.
Thus, we observe that the polarization  evidently originates from the finite
source effect.
\par
Using the above equations, we can express the polarization degree in the
bypass case as follows:
\ifReferee
\begin{eqnarray}
P(\rho_*|u)&\approx&
{7c_2\over
5(1-c_1/3)}{\left(\rho_*/u\right)^2\over\left(u^2+4\right)^2}
\nonumber\\&&\times
\left\{
1+\left[{304\over147}{u^6+3u^4+6u^2+5\over(u^2+2)(u^2+4)^2}-{60-28c_1\over15-5c_1}{u^2+1\over(u^2+2)(u^2+4)^2}
\right]\left({\rho_*\over u}\right)^2
\right\}.
\label{eq:pd}
\end{eqnarray}
\else
\begin{equation}
P(\rho_*|u)\approx
{7c_2\over
5(1-c_1/3)}{\left(\rho_*/u\right)^2\over\left(u^2+4\right)^2}
\left\{
1+\left[{304\over147}{u^6+3u^4+6u^2+5\over(u^2+2)(u^2+4)^2}-{60-28c_1\over15-5c_1}{u^2+1\over(u^2+2)(u^2+4)^2}
\right]\left({\rho_*\over u}\right)^2
\right\}.
\label{eq:pd}
\end{equation}
\fi
\begin{figure*}
\begin{center}
\rotatebox{-90}{\includegraphics[height=0.9\linewidth,clip]{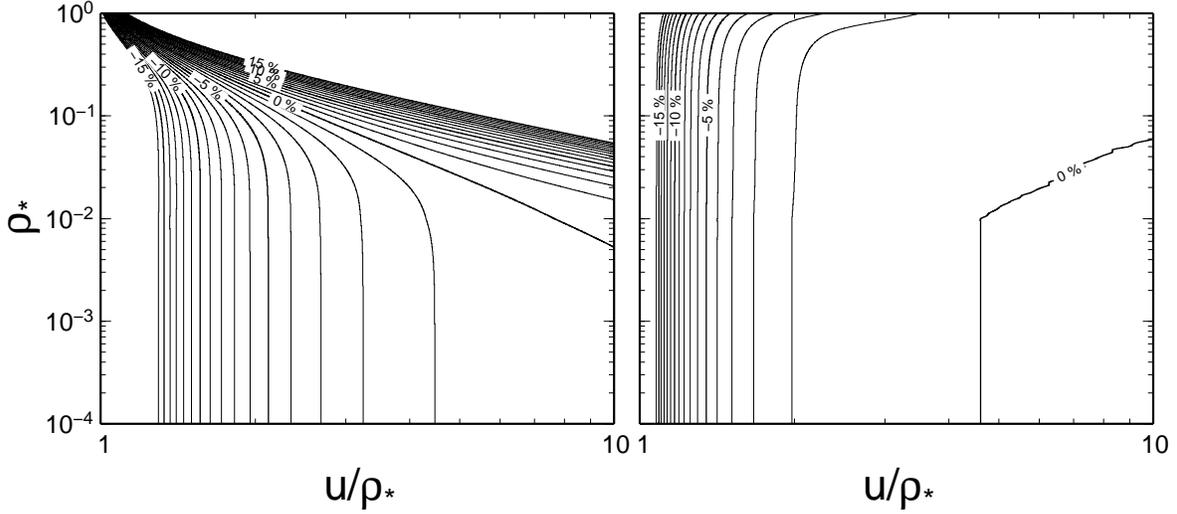}}
\caption{A Comparison of approximations to the  polarization
 degree: The left panel shows contours of the relative errors of an
 approximation given by \citeauthor{schn87} to the polarization degree
 estimated by using equations (\ref{eq:totFlux}) and (\ref{eq:ASU})
 numerically. The right panel shows contours of the relative errors of
 our  approximation (\ref{eq:pd}) to the numerical estimation. The
 contour spacing is $1~ \rmn{per~cent}$.
}%%% Figure 4
\label{fig:schvsus}
\end{center}
\end{figure*}
Equation (\ref{eq:pd}) is similar to the one derived by \citeauthor{schn87};
however, our approximation seems more useful even in the region
$u\ga1$. Fig. \ref{fig:schvsus} present a comparison of approximation
and theirs. Although both approximations
deviate from the full expression (\ref{eq:p_def}) with a factor $1.5-2$
at $u/\rho_*\sim1$, the relative error of our approximation is less than
$10~\rmn{per~cent}$ for $1.2\la u/\rho_*<10$ and $\rho_*<0.1$, 
whereas the region with the relative error less than $10~\rmn{per~cent}$
of the one given by \citeauthor{schn87} exists for $1.6\la u/\rho_*<10$
and $\rho_*<0.05$. 
\par
In the bypass case, the polarization degree has a maximum value
$P_\mathrm{max}$ at $\tau=0$, as follows:
\begin{equation}
P_\rmn{max}\approx{7c_2\over80(1-c_1/3)}\left({\rho_*\over
					    u_0}\right)^2
\left\{
1+\left[{871\over5880}+{1\over20(1-c_1/3)}
\right]\left({\rho_*\over u_0}\right)^2
\right\}, 
\label{eq:pmax}
\end{equation}
for $\rho_*\le u_0\ll 1$ and $1.2<u_0/\rho_*<5$. For $c_1=0.64$ and
$c_2=0.04$ (\citeauthor{schn87}), we obtain the maximum value of the polarization
degree, $P_\mathrm{max}\approx0.12(2\rho_*/u_0)^2~\rmn{per~cent}$. 
This value is very small, but not always undetectable. 
In fact, a polarization degree of this order was detected by
\citeauthor{kemp83}, who obtained the polarization degree
$\sim0.01~\rmn{per~cent}$ during
an eclipse in a binary system. 
\section{Extraction of the finite source effect from polarization}
In the previous section, we investigated the polarization by a
microlensing event. As mentioned in Introduction, combining the finite
source effect with the parallax effect enables us to decouple the
degeneracy in the microlens parameters. In this section,  we shall
consider the method of extracting information on the finite source
effect from the polarization. 
\subsection{Variability of the polarization degree}
\cite{simm95b} showed that the polarization degree varies with time by a
microlensing event and that the time profiles (variabilities) are
classified into two cases, i.e. transit case and bypass case. 
We thus discuss the finite source effect on the polarization degree
separately for each case.
\subsubsection{Transit case}
In the transit case, the polarization degree has a double-peak profile.
Therefore, we can measure the peak times $t_1, t_2 (>t_1)$ from the
polarimetric observation. 
According to \cite{schn87}, it is theoretically known that the
polarization degree has peaks near $u=0.96\rho_*$.
Thus, by combining $\Delta t=t_2-t_1$ with the analysis from the light
curve ($u_0, t_\rmn{E}$), we can estimate $\rho_*$ as follows:
\begin{equation}
\rho_*\approx1.04\sqrt{u_0^2+\left({\Delta t\over2t_\rmn{E}}\right)^2}.
\label{eq:tran_rho}
\end{equation}
\subsubsection{Bypass case}
Unfortunately, the polarization has only one
peak in the bypass case. Hence, we cannot measure $\Delta t$, which is
possible in the transit case. However, we may be able to use equation
(\ref{eq:pd}) to estimate $\rho_*$. 
Provided that $\rho_*$ has a small value and $1<u/\rho_*\la5$, 
we can approximate the equation as follows:
\begin{equation}
P={a\over u^2}\left(1+{b\over u^2}\right),\
 a={7c_2\rho_*^2\over80(1-c_1/3)},\ 
b=\left[{871\over5880}+{1\over20(1-c_1/3)}\right]\rho_*^2.
\end{equation}
Using the same approximation, we can deduce 
$A(\rho_*|u)/A_\mathrm{p}(u)$ from equation (\ref{eq:SI_lu0}) as
\begin{equation}
{A(\rho_*|u)\over A_\rmn{p}(u)}=1+\left[
{7\over40}-{1\over20(1-c_1/3)}
\right]\left({\rho_*\over u}\right)^2\equiv1+{c\over u^2},\ c=\left[
{7\over40}-{1\over20(1-c_1/3)}\right]\rho_*^2.
\end{equation}
Therefore, we can obtain the values of $a,b,c$ from the data fitting of the
light curve and the variability of the polarization degree in the same
event. Consequently, the coefficients $a,b,c$ yield  $\rho_*$, $c_1$ and $c_2$
as follows: 
\begin{equation}
\rho_*=\sqrt{{294\over95}(b+c)},\ c_1={15(147b-233c)\over1029b-871c}\
 \mbox{and}\ c_2={7600a\over7(1029b-871c)}.
\end{equation}
\subsection{$\bmath{u/\brho_*}$ dependence of polarization for small $\bmath{\brho_*,  u}$}
As mentioned in subsection 3.2, for small $\rho_*$ and $u$, the
polarization degree depends only on $u/\rho_*$. From equations
(\ref{eq:totFlux}) and (\ref{eq:ASU}), as well, we can prove the
validity of this statement. In these equations, the arguments of the
complete elliptical integrals, $k$ and $n$ are approximated as $k^2\approx
n=4t/(1+t)^2$ for $\rho, u\ll1$ (but $0<t=\rho/u<\infty$). By using the
relation $(u-\rho)^2\Pi(\pi/2;k,n)\approx (\rho+u)^2E(\pi/2;k)$ derived
from equation (\ref{eq:A9}) in Appendix A, we find that $F(\pi/2;k)$
terms are the leading terms in the integrands of equations
(\ref{eq:totFlux}) and (\ref{eq:ASU}). 
Therefore, we can approximate these equations as follows:
\begin{eqnarray}
S_I(\rho_*|u)&\approx&{4S_{I0}\over\pi(1-c_1/3)}{u\over\rho_*^2}
\int_0^{\rho_*/u}\rmn{d}t\left\{
1-c_1\left[
1-\sqrt{1-(ut/\rho_*)^2}\right]\right\}
{tF(\pi/2;k(t))\over1+t},\label{eq:Ftotap}\\
\noalign{\hbox{\mbox{and}}}
D(\rho_*|u)&\approx&{4c_2S_{I0}\over\pi(1-c_1/3)}{u\over\rho_*^2}
\int_0^{\rho_*/u}\rmn{d}t\left[1-\sqrt{1-(ut/\rho_*)^2}\right]
{(3t^4+2t^2+3)F(\pi/2;k(t))\over8t(1+t)}.\label{eq:Dap}
\end{eqnarray}
From these equations, we have proved that
$P(\rho_*|u)=D(\rho_*|u)/S_I(\rho_*|u)$ depends only on $u/\rho_*$ for
$\rho_*,u\ll1$. 
\begin{figure*}
\begin{center}
\rotatebox{-90}{\includegraphics[width=0.4\linewidth]{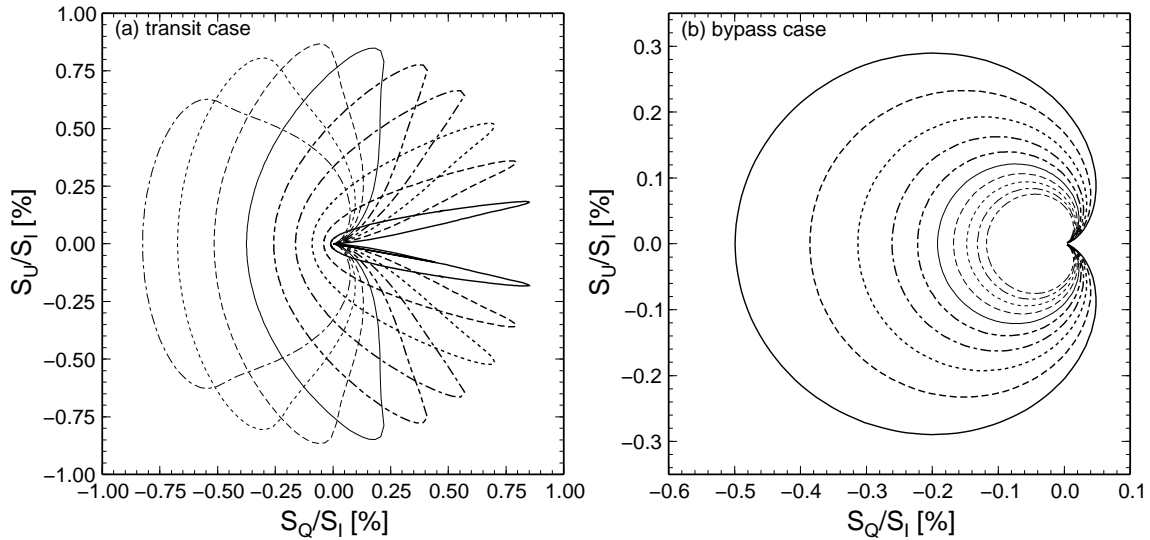}}
\caption{$QU$ plots in transit cases (a) and in bypass
 cases (b): In both cases the source size $\rho_*=10^{-3}$ and
 the direction of the relative velocity $\bmath{V}$ is assumed to be
 parallel to the observer's  $x$-axis ($\alpha=0$). In the left panel,
 the ratios of $u_0$ to $\rho_*$ are 0.9, 0.8, 0.7, 0.6, 0.5, 0.4, 0.3,
 0.2 and 0.1 from the leftmost curve. In the right panel, they are
 obtained as 1.1, 1.2, 1.3, 1.4, 1.5, 1.6, 1.7, 1.8, 1.9 and 2.0 from
 the outer curve. 
}%%% Figure 5
\label{fig:quplot}
\end{center}
\end{figure*}
\par
Hence, we define the normalized Stokes parameters $\widehat{S}_Q$ and 
$\widehat{S}_U$ as 
\begin{equation}
\widehat{S}_Q=S_Q/S_I=P(\rho_*|u)\cos2(\vartheta+\alpha)\ \mbox{and}\
\widehat{S}_U=S_U/S_I=-P(\rho_*|u)\sin2(\vartheta+\alpha),
\label{eq:normStokes}
\end{equation}
respectively, where $\vartheta=-\tan^{-1}u_0/\tau$ and $\alpha$ is the
direction of $\bmath{V}$ to the $x$-axis. Then, we can draw $QU$ plots for
some values of $u_0/\rho_*$. We have presented various $QU$ plots for
various $u_0/\rho_*$ values in Fig. \ref{fig:quplot}, where for all the
curves, we employ the same values of $\rho_*$ and $\alpha$;
$\rho_*=0.001$ and $\alpha=0$. These curves are very 
different from each other. In particular, the curves in the transit case
(panel (a)) differ considerably from those in the bypass case (panel
(b)). As long as $\rho_*\ll1$, however, we can specify the shapes of these
curves by one parameter, $u_0/\rho_*$, because the polarization degree
depends only on the parameter $u/\rho_*$ for small $\rho_*$. Hence,
if we can fit the data of the Stokes parameters on an appropriately
scaled curve, we can estimate $u_0/\rho_*$.  
\par
\begin{figure*}
\begin{center}
\rotatebox{-90}{\includegraphics[width=0.6\linewidth]{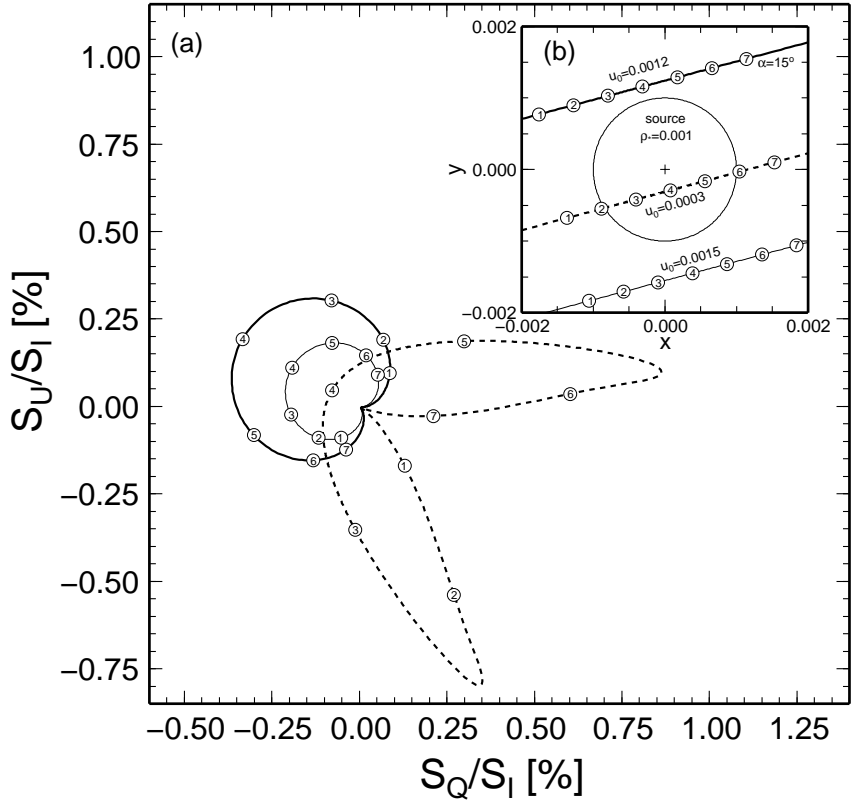}}
\caption{$QU$ plots and the corresponding trajectories of the
 microlenses: When the microlens moves from 1 to 7 on each line in panel
 (b), the point on the $QU$ plot moves from 1 to 7 on the corresponding
 curve in panel (a). In this figure, it is assumed that $\rho_*=10^{-3}$
 and the lines are inclined to the $x$-axis with $\alpha=15^\circ$. The
 time interval between the circled points is $5\times10^{-4}t_\rmn{E}$. 
}%%% Figure 6
\label{fig:qutraj}
\end{center}
\end{figure*}

Besides the possibility of estimating $\rho_*$, we should note that
the $QU$ plot provides information on the direction of the relative
microlens motion $\bmath{V}$. Microlensed $QU$ plots are shown in
Fig. \ref{fig:qutraj} (panel (a)), in the cases when the microlenses
move along lines inclined to the observer's $x$-axis with an angle
$\alpha=15^\circ$ (panel (b)). In these cases, we find that the plots
lean at an angle $-2\alpha$ as compared to the ones with $\alpha=0$ (as
shown in Fig. \ref{fig:quplot}). In Fig. \ref{fig:qutraj}, we draw the
slanted $QU$ plots for three cases: 
when the microlens moves from the lower-left corner to the right in
panel (b) in the bypass case ($u_0=0.0015$), the same movement in the
transit case ($u_0=0.0003$) and when it moves from the left to
the upper-right corner in the bypass case ($u_0=0.0012$). While, in the first
two cases, a point on the $QU$ plot moves clockwise around the origin
with time, in the last case, the point moves anti-clockwise. Hence, we
find that points on the $QU$ plots, in general, move anti-clockwise
when the vector $\bmath{u}\times\bmath{V}$ points to the observer from
the lens plane, and vice versa. 
\par
Thus, if we have data on the Stokes parameters, i.e. $(S_I, S_Q, S_U)$
during a microlensing event, we can obtain not only the finite source
effect $\rho_*$ but also the direction of the relative microlens motion,
including the sign of $\bmath{u}\times\bmath{V}$.
\section{Summary and Discussion}
In this paper we have investigated the single microlens effect on the
Stokes parameters in order to obtain some semi-analytical formulae.
As a result, we observe that the formulae not only reduce the double
integrals in the estimations of the parameters but also can be
expressed as a useful approximation form in the bypass case. In
addition, we have shown that by combining the polarimetric data with
photometric data, we can estimate not only the finite source effect
$\rho_*$  but also the direction of the microlens motion, including the
sign of $\bmath{u}\times\bmath{V}$.
\par
Although we have argued on the benefits of observations of polarized stars
by the microlensing event, we have not yet enumerated the difficulties
involved in the polarimetric observation of the microlensed star. In
this regard, we shall comment on the following difficulties in the
remainder of this paper: 
(1) the lower limit of the detectable polarization degree, 
(2) the limiting magnitude of the polarimetric observation,
(3) the observable duration for polarization
and
(4) the identification of the target star.
\par
The first difficulty is concerned with observational
technology. Although \cite{kemp83} succeeded in measuring the
polarization degree with $\sim0.01~\rmn{per~cent}$ accuracy during an
eclipse in a binary system, it is not easy to detect polarized light
with such accuracy even with the present technology. At present, we can
expect the accuracy of the polarization degree to be
$P_\rmn{lim}\approx0.05~\rmn{per~cent}$ (Hirata, in private
communication). This limit defines the upper limit for $u$ as
$u\la3\rho_*$ for $\rho_*<0.1$ (see
Fig. \ref{fig:polarization2}). Hence, we can define the effective radius  
$R$ in the polarimetric observation of a microlensed star that depends
on $P_\rmn{lim}$, e.g. $R\approx3\rho_*$ for
$P_\rmn{lim}=0.05~\rmn{per~cent}$.
\footnote{%%% footnote 4 
The effective radius, of course, depends on coefficients $c_1, c_2$ in
equations (\ref{eq:surfbr}) and (\ref{eq:IQUV}). The value estimated
here is in the case where $c_1=0.64$ and $c_2=0.04$ \citep{schn87}.
}
\par
The second difficulty arises due to the fact that the limiting magnitude
of polarimetry is in general $5-7$ mag brighter than that of photometric
observation. Most of microlensed stars have magnitudes in the range of
$15-21$ mag in the photometry ($I$-band). This means that the
corresponding magnitudes in the polarimetry of the star are in the range
of $20-28$ mag. Therefore, in order to detect the polarization, we need
a large telescope. In FOCAS of the Subaru Telescope, the limiting
magnitude is 26.2 mag in the $I$-band with a 1200 s exposure
\citep{Kashi02}. Thus, the observation is very difficult even by using
FOCAS. Fortunately, while the polarization is detectable, the
magnification factor of the star is larger than $1/R$. Hence, 
the polarized light from the microlensed star with even the faintest
magnitude may be detectable by FOCAS because the magnification factor is
larger than 10 when $\rho_*<0.03$.
\par
The third difficulty is associated with the exposure time for the
microlensed star. For the observation of faint stars, a long exposure
time is necessary. In order to extract the finite source effect, we need
to observe the lensed star a number of times. Further, the total
exposure time must be less than the observable duration for
polarization, which is equal to the crossing time $\Delta t_\rmn{c}$ of
the microlens over the effective diameter ($=2R$),
i.e. $2Rt_\rmn{E}\approx6\rho_*t_\rmn{E}$. We thus obtain the lower 
limit of $\rho_*$ in the polarimetric observation of the microlensed
star as follows: 
\begin{equation}
4.6\times10^{-3}\left({N_\rmn{obs}\over100}\right)
\left({t_\rmn{exp}\over1200~\rmn{s}}\right)
\left({50~\rmn{days}\over t_\rmn{E}}\right)<\rho_*,
\label{eq:lower_rho}
\end{equation}
where $N_\rmn{obs}$ and $t_\rmn{exp}$ denote the frequency of the
observations and the exposure time, respectively.
\par
Among the four difficulties listed above, the last one may be the most
challenging. It arises from a reason specific to the microlensing
event. Generally, microlensed stars are observed in very crowded
regions such as our Galactic bulge, LMC and SMC. Hence, there are many
stars around the target star within the same field. As $50-100$ stars
can be simultaneously observed using FOCAS (Kawabata, in private
communication), it may be possible to detect the polarization.
\par
If we can overcome the above difficulties, we can estimate the finite
source effect from the polarimetric observation and decouple the
degeneracy in the microlens parameters by using the relation given by
\cite{vanBell99}. Even after satellite projects succeed in
measuring the angular Einstein ring radius $\theta_\rmn{E}$,
the polarimetric observation of the microlensed star will provide useful
information such as on the limb-darkening model, on the relation between
angular diameter and colour-magnitude of stars (\citeauthor{vanBell99})
and so on. We thus believe that it is important and worthwhile to focus
attention on polarimetry in microlensing events after receiving
microlens alerts such as from the Early Warning System \citep{Udal03}.
\footnote{%%% footnote 5
\texttt{http://www.astrouw.edu.pl/$^\sim$ogle/ogle3/ews/ews.html}}

\appendix
\section{Elliptical Integrals}
The elliptical functions of the first, second and third kind ($F,E,\Pi$)
are given as follows:
 \begin{eqnarray}
F(\phi;k)&=&\int_0^\phi{\rmn{d}\theta\over\sqrt{1-k^2\sin^2\theta}}
=\int_0^{\sin^{-1}\phi}{\rmn{d}t\over\sqrt{(1-t^2)(1-k^2t^2)}}, 
\label{eq:cel1}\\
E(\phi;k)&=&\int_0^\phi\rmn{d}\theta\sqrt{1-k^2\sin^2\theta}
=\int_0^{\sin^{-1}\phi}dt\sqrt{1-k^2t^2\over1-t^2},\\
\label{eq:cel2}
\Pi(\phi;k,n)&=&\int_0^\phi{\rmn{d}\theta\over(1-n\sin^2\theta)
\sqrt{1-k^2\sin^2\theta}}
=\int_0^{\sin^{-1}\phi}{\rmn{d}t\over(1-nt^2)\sqrt{(1-t^2)(1-k^2t^2)}}.
\label{eq:cel3}
\end{eqnarray}
Let us define the integration as follows:
\begin{equation}
\Lambda_l(t;k,n)
\equiv\int_0^t{\rmn{d}t\over(1-nt^2)^l\sqrt{(1-t^2)(1-k^2t^2)}}
\end{equation}
The objective in this appendix is to express $\Lambda_l$ as a
combination of $F,E$ and $\Pi$. First, we consider the derivative of
$[t\sqrt{f(t)}]/(1-nt^2)^{l-1}$ with respect to $t$, where
$f(t)=(1-t^2)(1-k^2t^2)$:
\begin{equation}
{\rmn{d}\ \over \rmn{d}t}\left[{t\over(1-nt^2)^{l-1}}\sqrt{f(t)}\right]
= {2(l-1)f(t)+(1/2)\left(1-nt^2\right)
\Bigl\{
tf^\prime(t)+2(3-2l)f(t)
\Bigr\}
\over\sqrt{f(t)}(1-nt^2)^l},\label{eqn:der}
\end{equation}
We restrict ourselves to $l\ge2$. From the definition of $\Lambda_l$,
the first three functions, $\Lambda_1,\Lambda_0, \Lambda_{-1}$, are
expressed by the following three complete elliptical functions,
\ifReferee
\begin{eqnarray}
&&\Lambda_1(t;k,n)=\Pi(\sin^{-1}t;k,n),\nonumber\\
&&\Lambda_0(t;k,n)=F(\sin^{-1}t;k), \nonumber\\
\noalign{\hbox{\mbox{and}}}
&&\Lambda_{-1}(t;k,n)=(1-\displaystyle{n\over k^2})F(\sin^{-1}t;k)
 +\displaystyle{n\over k^2}E(\sin^{-1}t;k),\nonumber
\end{eqnarray}
\else
\begin{equation}
\Lambda_1(t;k,n)=\Pi(\sin^{-1}t;k,n),\quad
\Lambda_0(t;k,n)=F(\sin^{-1}t;k)\quad \mbox{and}\quad
\Lambda_{-1}(t;k,n)=(1-\displaystyle{n\over k^2})F(\sin^{-1}t;k)
 +\displaystyle{n\over k^2}E(\sin^{-1}t;k),
\end{equation}
\fi
respectively. We denote the numerator of equation (\ref{eqn:der}) as
$g(t)$
\begin{equation}
g(t)=2(l-1)f(t)+{1\over2}(1-nt^2)
\left[tf^\prime(t)-2(2l-3)f(t)\right]
=A+B(1-nt^2)+C(1-nt^2)^2+D(1-nt^2)^3,\nonumber
\end{equation}
where 
\ifReferee
\begin{eqnarray}
A=2(l-1)\left(1-{1\over n}\right)\left(1-{k^2\over n}\right),&&
\hspace{-5mm}
B=-(2l-3)\left(1-2{1+k^2\over n}+3{k^2\over n^2}\right),\nonumber\\
C=2(l-2)\left(3{k^2\over n^2}-{1+k^2\over n}\right)&&
\hspace{-5mm}
\mbox{and}\quad
D=-(2l-5){k^2\over n^2}.\nonumber
\end{eqnarray}
\else
\begin{eqnarray}
A=2(l-1)\left(1-{1\over n}\right)\left(1-{k^2\over n}\right),\
B=-(2l-3)\left(1-2{1+k^2\over n}+3{k^2\over n^2}\right),\
C=2(l-2)\left(3{k^2\over n^2}-{1+k^2\over n}\right)\ \mbox{and}\
D=-(2l-5){k^2\over n^2}.\nonumber
\end{eqnarray}
\fi
These coefficients enable us rewrite equation (\ref{eqn:der}) as follows:
\begin{eqnarray}
{\rmn{d}\ \over \rmn{d}t}\left[{t\over(1-nt^2)^{l-1}}\sqrt{f(t)}\right]&=&
  A{\rmn{d}\Lambda_l\over \rmn{d}t} 
+B{\rmn{d}\Lambda_{l-1}\over \rmn{d}t}
+C{\rmn{d}\Lambda_{l-2}\over \rmn{d}t}
+D{\rmn{d}\Lambda_{l-3}\over \rmn{d}t},\nonumber\\
&=&2(l-1)\left(1-{1\over n}\right)
\left(1-{k^2\over n}\right){\rmn{d}\Lambda_l\over \rmn{d}t}
-(2l-3)\left(1-2{1+k^2\over n}+3{k^2\over n^2}\right)
{\rmn{d}\Lambda_{l-1}\over \rmn{d}t}
\nonumber\\&&
+2(l-2)\left(3{k^2\over n^2}-{1+k^2\over n}\right)
{\rmn{d}\Lambda_{l-2}\over \rmn{d}t}
-(2l-5){k^2\over n^2}{\rmn{d}\Lambda_{l-3}\over \rmn{d}t},
\end{eqnarray}
We integrate the above equation from $t=0$ to $1$ in order to obtain the
following recursion:
\ifReferee
\begin{eqnarray}
2(l-1)\left(1-{1\over n}\right)\left(1-{k^2\over n}\right)\Lambda_l(1;k,n)
&=&
(2l-3)\left(1-2{1+k^2\over n}+3{k^2\over n^2}\right)\Lambda_{l-1}(1;k,n)
\nonumber\\
&&
 -2(l-2)\left(3{k^2\over n^2}-{1+k^2\over n}\right)\Lambda_{l-2}(1;k,n)
\nonumber\\
&&+(2l-5){k^2\over n^2}\Lambda_{l-3}(1;k,n). \label{eqn:A7}
\end{eqnarray}
\else
\begin{eqnarray}
2(l-1)\left(1-{1\over n}\right)\left(1-{k^2\over n}\right)\Lambda_l(1;k,n)
&=&
(2l-3)\left(1-2{1+k^2\over n}+3{k^2\over n^2}\right)\Lambda_{l-1}(1;k,n)
\nonumber\\
&&
 -2(l-2)\left(3{k^2\over n^2}-{1+k^2\over n}\right)\Lambda_{l-2}(1;k,n)
%\nonumber\\&&
+(2l-5){k^2\over n^2}\Lambda_{l-3}(1;k,n). \label{eqn:A7}
\end{eqnarray}
\fi
For example, in the cases  of $l=2,3$, we find 
\ifReferee
\begin{eqnarray}
2\left(1-{1\over n}\right)\left(1-{k^2\over n}\right)\Lambda_2(1;k,n)
&=&\left(1-2{1+k^2\over n}+{3k^2\over n^2}\right)\Pi(\pi/2;k,n)\nonumber\\
&&+{1\over n}\left(1-{k^2\over n}\right)F(\pi/2;k)-{1\over n}E(\pi/2;k),
\label{eq:A9}
\end{eqnarray}
\else
\begin{equation}
2\left(1-{1\over n}\right)\left(1-{k^2\over n}\right)\Lambda_2(1;k,n)
=\left(1-2{1+k^2\over n}+{3k^2\over n^2}\right)\Pi(\pi/2;k,n)
+{1\over n}\left(1-{k^2\over n}\right)F(\pi/2;k)-{1\over n}E(\pi/2;k),
\label{eq:A9}
\end{equation}
\fi
and
\ifReferee
\begin{eqnarray}
4\left(1-{1\over n}\right)\left(1-{k^2\over n}\right)\Lambda_3(1;k,n)
=
&&3\left(1-2{1+k^2\over n}+{3k^2\over n^2}\right)\Lambda_2(1;k,n)
\nonumber\\&&
 -2\left(3{k^2\over n^2}-{1+k^2\over n}\right)\Pi(\pi/2;k,n)
+{k^2\over n^2}F(\pi/2;k).
\end{eqnarray}
\else
\begin{eqnarray}
4\left(1-{1\over n}\right)\left(1-{k^2\over n}\right)\Lambda_3(1;k,n)
=
3\left(1-2{1+k^2\over n}+{3k^2\over n^2}\right)\Lambda_2(1;k,n)
 -2\left(3{k^2\over n^2}-{1+k^2\over n}\right)\Pi(\pi/2;k,n)
+{k^2\over n^2}F(\pi/2;k).
\end{eqnarray}
\fi
We substitute $n=4\rho u/(u+\rho)^2, k^2=4n/[(u-\rho)^2+4]$ into
equation (\ref{eqn:A7}) 
to obtain a recursive formula as follows:
\ifReferee
\begin{eqnarray}
\Lambda_l(1;k,n)&=&\left({2l-3\over l-1}\right)
{\rho^2+u^2-2\over (u-\rho)^2}\Lambda_{l-1}(1;k,n)%
-\left({l-2\over l-1}\right)
{(\rho^2-u^2)^2-8(\rho^2+u^2)\over (\rho-u)^4}
\Lambda_{l-2}(1;k,n)%
\nonumber\\&&\hspace{0cm} 
-2\left({2l-5\over l-1}\right)
{(\rho+u)^2\over(\rho-u)^4}\Lambda_{l-3}(1;k,n).
\end{eqnarray}
\else
\begin{eqnarray}
\Lambda_l(1;k,n)&=&\left({2l-3\over l-1}\right)
{\rho^2+u^2-2\over (u-\rho)^2}\Lambda_{l-1}(1;k,n)%
-\left({l-2\over l-1}\right)
{(\rho^2-u^2)^2-8(\rho^2+u^2)\over (\rho-u)^4}
\Lambda_{l-2}(1;k,n)%
\nonumber\\&&\hspace{49mm}
-2\left({2l-5\over l-1}\right)
{(\rho+u)^2\over(\rho-u)^4}\Lambda_{l-3}(1;k,n).
\end{eqnarray}
\fi
For example, in the cases of $l=2,3$, we obtain 
\ifReferee
\begin{eqnarray}
\Lambda_2(1;k,n)&=&{u^2_0+\rho^2-2\over(u-\rho)^2}\Pi(\pi/2;k,n)
-{1\over2}\left({u+\rho\over u-\rho}\right)^2F(\pi/2;k)
\nonumber\\&&
+{(u+\rho)^2\left\{(u-\rho)^2+4\right\}\over2(u-\rho)^4}E(\pi/2;k),\\
\Lambda_3(1;k,n)&=&{u^4+4u^2\rho^2-2(u^2+\rho^2)
+\rho^4+6\over(u-\rho)^4}\Pi(\pi/2;k,n)
-{(u+\rho)^2(3u^2+3\rho^2-2)\over4(u-\rho)^4}F(\pi/2;k)
\nonumber\\&&
   +{3(u+\rho)^2(u^2+\rho^2-2)\left\{(u-\rho)^2+4\right\}\over4(u-\rho)^6}
    E(\pi/2;k).
\end{eqnarray}
\else
\begin{eqnarray}
\Lambda_2(1;k,n)&=&{u^2_0+\rho^2-2\over(u-\rho)^2}\Pi(\pi/2;k,n)
-{1\over2}\left({u+\rho\over u-\rho}\right)^2F(\pi/2;k)
+{(u+\rho)^2\left\{(u-\rho)^2+4\right\}\over2(u-\rho)^4}E(\pi/2;k),\\
\Lambda_3(1;k,n)&=&{u^4+4u^2\rho^2-2(u^2+\rho^2)
+\rho^4+6\over(u-\rho)^4}\Pi(\pi/2;k,n)
-{(u+\rho)^2(3u^2+3\rho^2-2)\over4(u-\rho)^4}F(\pi/2;k)
\nonumber\\&&
   +{3(u+\rho)^2(u^2+\rho^2-2)\left\{(u-\rho)^2+4\right\}\over4(u-\rho)^6}
E(\pi/2;k).
\end{eqnarray}
\fi
\end{document}